\documentclass{iauc}
\usepackage{graphicx}

\title[Near-IR L-Z relation] %% give here short title %% 
{The near-IR luminosity-metallicity relation of dwarf irregular galaxies}

\author[I. Saviane et al.]   %% give here short author list %%
{Ivo Saviane$^1$,   % 1-ESO
Enrico V. Held$^2$,   % 2-(OAPD)
Valentin Ivanov$^1$,  % 1-(ESO)
Danielle Alloin$^3$,  % 3-Saclay
Fabio Bresolin$^4$,   % 4-(IfA Hawaii)
R. Michael Rich$^5$,  % 5-(UCLA)
Luca Rizzi$^4$ \and   % 4-(IfA Hawaii)
Yazan Momany$^6$      % 6-(UniPD)      
}

\affiliation{$^1$ ESO,
$^2$ OAPD,       %\\[\affilskip]
$^3$ CEA,     %\\[\affilskip]
$^4$ IfA Hawaii, %\\[\affilskip]
$^5$ UCLA,      %\\[\affilskip]
$^6$ UniPD      %\\[\affilskip]
}

\pubyear{2005}
\volume{198} %% insert here IAU Symposium No.
\pagerange{xxx -- xxx}
\date{?? and in revised form ??}
\jname{ Near-Field Cosmology With Dwarf Elliptical Galaxies}
\editors{B. Binggeli, and H. Jerjen, eds.}

\def\lz{L-Z}
\newcommand\ion[2]{#1$\;${\small\rm{#2}}\relax} % from emulateapj 

\begin{document}

\maketitle

\begin{abstract}
We report on the recent developments of our long-term investigation of
the near-IR luminosity-metallicity relation for dwarf irregular
galaxies in nearby groups. A very well-defined relation is emerging from
our observational database, and a preliminary discussion of its
implications is given.

\end{abstract}

\firstsection % if your document starts with a section,

\section{Introduction}

Since a few years, we are carrying out a program to collect near-IR
imaging of dwarf irregular (dIrr) galaxies in galaxy groups, and optical
spectra of their H{\sc ii} regions, to test the existence of a
luminosity-metallicity (\lz) relation.
The main purpose of the study is to provide a sound starting point to
discuss how the relation is created, which gives insight into the
interplay between processes like mass loss on galactic scales, and
galactic chemical evolution. In turn, this has implications on the
identification of the sources of metals in the intergalactic medium
(IGM), and on the cosmic chemical evolution.
The existence of a \lz\ relation is in fact
usually explained, at least in the case of spheroids,
as the result of mass loss through galactic winds triggered by SN
explosions (Larson \cite{larson74}).  Notwithstanding its importance,
the existence of a
\lz\ relation for dIrr is controversial, with some studies finding very well
defined correlations (Skillman et al. \cite{skh89}; Richer \& McCall
\cite{rm95}; Pilyugin \cite{pilyugin01b}), some finding mild relations with substantial scatter
(Skillman et al. \cite{skillman_scl_hii}), and some finding no
correlation at all (Hidalgo-G\'amez \& Olofsson \cite{anamaria}; Hunter \&
Hoffman \cite{hh99}).
So we decided to improve the situation by devising an observational
campaign to overcome some of the existing limitations.
The idea is 
(a)  to put
together homogeneous samples of oxygen abundances;  to consider only
galaxies in (b) well-defined environments, and (c) with reduced distance
ranges; and finally (d) to image galaxies in the near-IR. 
The campaign started as a pilot program at 
4-m class telescopes,
so it had been
restricted to the three nearest groups of
galaxies, namely M81, Centaurus~A, and Sculptor, whose barycenters lie
at similar distances ($\sim 3 \rm Mpc$) from the Sun. They also let us
probe diverse environments, since members of the Cen~A and M81 groups
show more interactions than those of the Scl group.
A progress report based on data for the Sculptor group has been
presented in Saviane et al. (\cite{sidney}). More recently we have been
able to obtain and reduce data for a few galaxies in the northern M81
group, so here we present an update of the project including data
for three more galaxies.

\section{The data sets}

\begin{figure}
\begin{tabular}{ccc}
 \includegraphics[width=4.3cm]{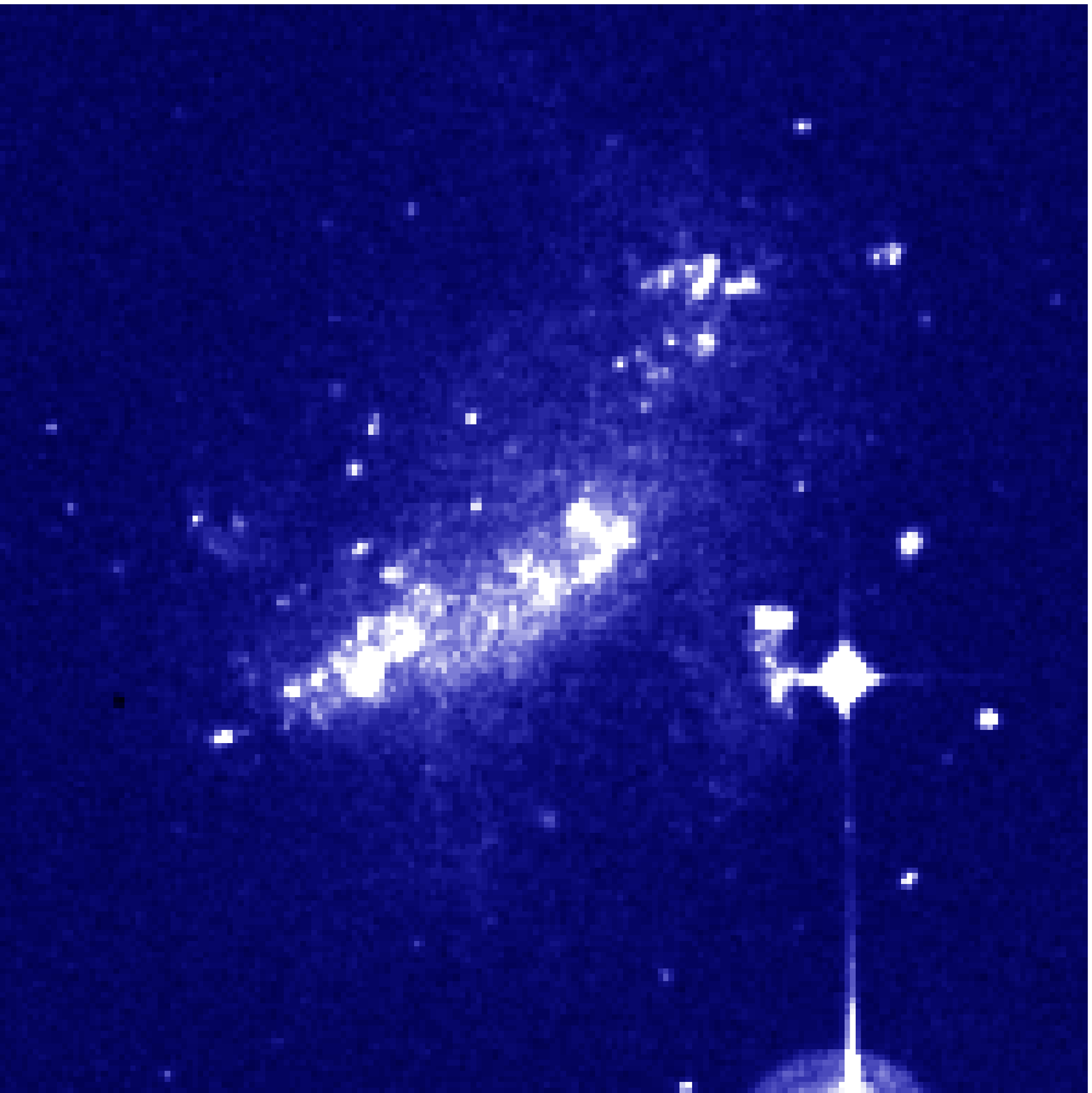}         &
 \includegraphics[width=4.3cm]{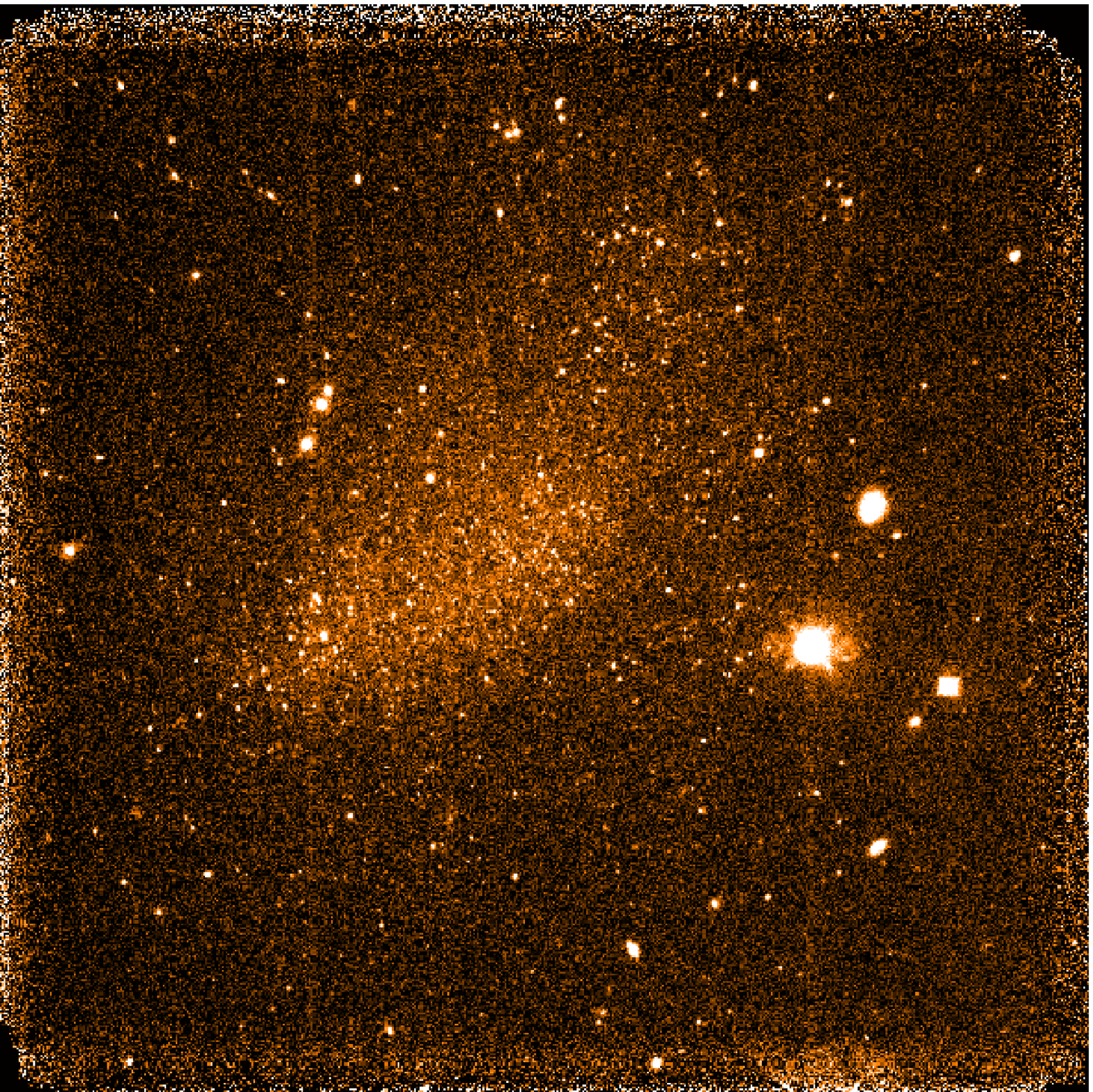}        &
 \includegraphics[width=4.3cm]{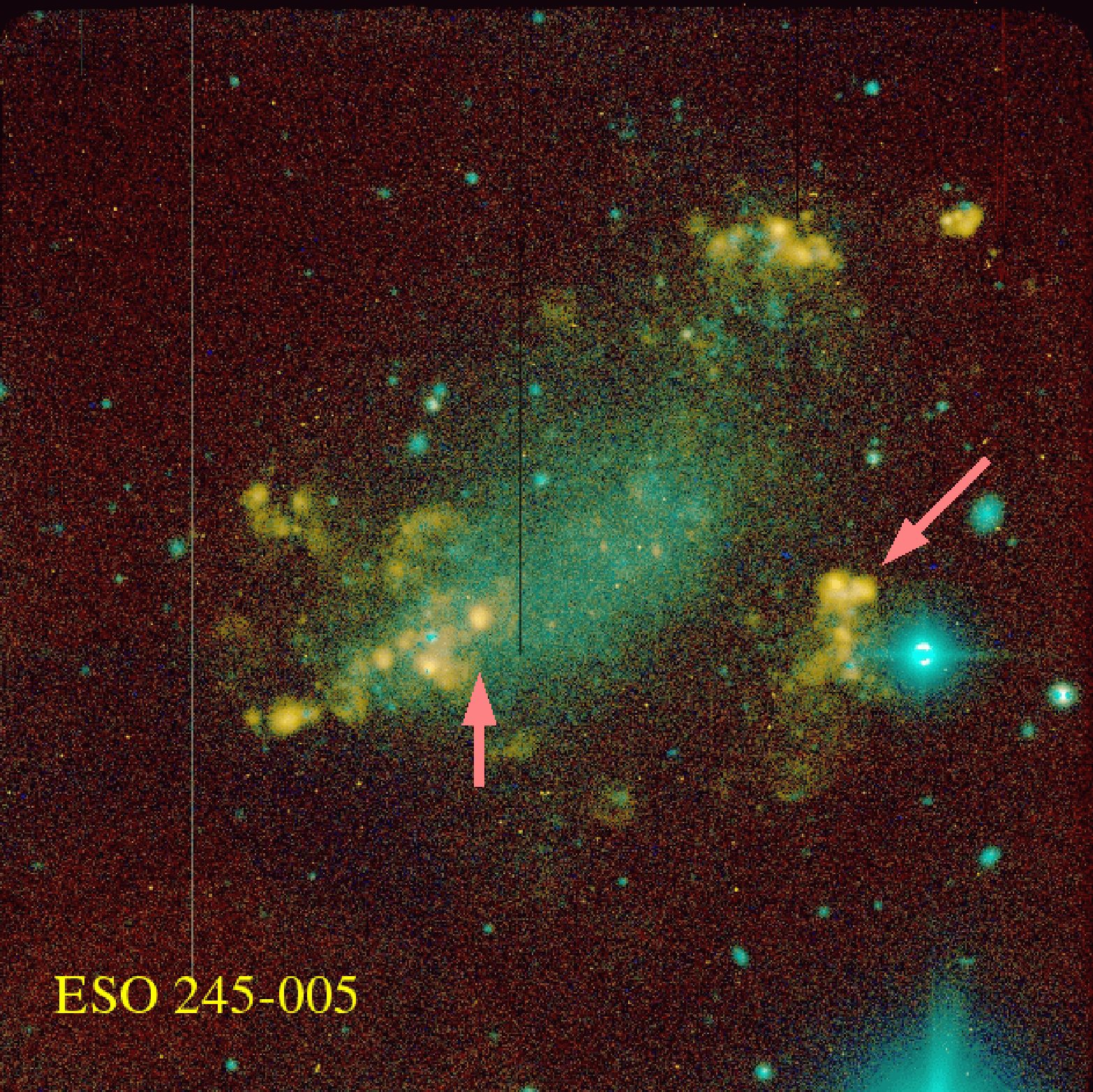}
\end{tabular}
  \caption{From left to right, we show the morphology of
  one of our targets (ESO~245-005) in the optical (DSS), infrared $H$
(SOFI), and H$\alpha$+R (EFOSC). The reddish, diffused nebulae in the
rightmost image are H{\sc ii} regions, and the two arrows mark those for which
spectra have been obtained. The images cover $\sim 5\times5$ arcmin$^2$.
\label{fig:eso245}}
\end{figure}

Nine galaxies of the Sculptor group have been observed so far, mostly in
the year 2002; ESO~245-005 was re-observed in October 2003, to get a
spectrum of a region closer to the optical center of the galaxy (see
Fig.~\ref{fig:eso245}). All data have been obtained at the La Silla
observatory using SOFI at the ESO/NTT and EFOSC at the ESO/3.6m. Typical
exposure times were of 1h for  spectroscopy and 1~hr on target +
1~hr on sky for the near-IR imaging in the $H$ band.
For the M81 group, ten galaxies have been observed in the course of the
years 2001 to 2003: interestingly, while only one galaxy of the Scl
group did not show H$\alpha$ emission, in this case $50\%$ of them did
not show any emission. For this northern group, the KAST spectrograph at
the Lick/3m telescope, and the INGRID camera at the ING/WHT telescope in
La Palma were used, with exposure times comparable to those listed
above.
If the suppressed SF activity (compared to the Scl group) is
confirmed, it will be telling us something about the effect of an
environment that is more 'risky'  for dwarf galaxies.
The reductions and calibrations of the data are thoroughly described in
Saviane et al. (2005, in preparation), and 
it is worth mentioning that, for the galaxies
in common, our abundances are consistent with those of Skillman et
al. (\cite{skillman_scl_hii}; see also Saviane et al. \cite{sidney}).

\section{The near-IR luminosity-metallicity relation}

\begin{figure}
\centerline{
 \includegraphics[width=10.5cm]{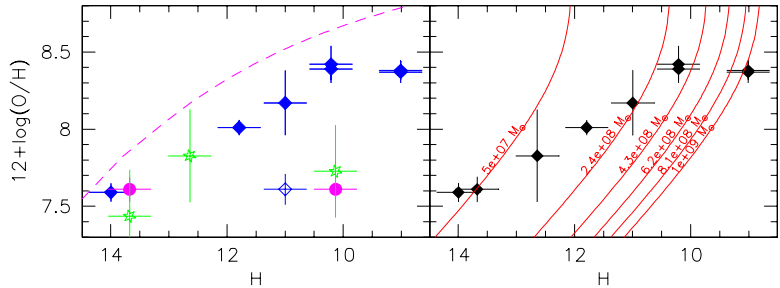} } \caption{ 
Metallicity of H{\sc ii} regions vs. apparent $H$ luminosity of galaxies in the
Scl and M81 groups. In the left panel, the diamonds represent H{\sc ii}
regions of the Scl group galaxies: filled diamonds are regions near the
center of each galaxy, while the open diamond is the external region of
ESO~245-005 (see also Fig.~\ref{fig:eso245}). M81 galaxies are
represented by filled circles (abundances obtained through the direct
method) and stars (abundances measured with the P00 method).  The dashed
line marks the metallicity reached if the chemical evolution is
truncated when the remaining gas mass goes below $2\times 10^7~m_\odot$.
In the right panel, our best data are compared with evolutionary tracks
for closed-box models with reduced yields (see text).
\label{fig:lz}}
\end{figure}

To construct our near-IR luminosity-metallicity relation, ideally one
would like to include only those galaxies for which abundances based on
the direct method could be obtained. This would mean only two galaxies
for the M81 group, thus, to improve the situation, we measured oxygen
abundances also using the indirect method proposed by Pilyugin
(\cite{pilyugin00}, P00), which allows to include all three galaxies for
which reliable spectra could be obtained.

Oxygen abundances vs. apparent $H$ magnitudes are presented in
Fig.~\ref{fig:lz}: the left panel shows the whole data set, while the
right panel shows a fiducial subsample.
Since the galaxies are at a similar average distance, we do not need to
correct for distance modulus: in any case the horizontal error bars
show the variation in luminosity due to a $1$~Mpc distance uncertainty
at $3$~Mpc. The external region of ESO~245-005 (at $H=11$) is
under-abundant by $\sim 0.5$~dex compared to the central one. The region
we measured in DDO~42 (near $H=10$) is also external, but we cannot make
a comparison with a central one. Finally, metallicities estimated with
the P00 method differ by at most $\pm 0.2$~dex from those
measured using the direct method, in agreement with the findings
of Skillman et al. (\cite{skillman_scl_hii}).
In the right panel of the figure we plot only the central H{\sc ii} regions,
including the one having the abundance estimated with the P00 method, to
which we assign a $50\%$ error bar. With this selection, a clear
relation emerges, in the usual sense of having higher oxygen abundances
for more massive galaxies. Our strategy has allowed
to construct a relation that is much better defined than the existing
ones.  It also seems like the dependence on luminosity 'saturates' when
the brightest galaxy in our sample is reached (NGC~625): excluding this
galaxy, a linear regression has a correlation coefficient $r=-0.996$,
and a slope of $-0.22\pm 0.01$~dex/mag.

The interpretation of the relation is not straightforward, though. As it
was recalled in the Introduction, in the case of dE/dSph galaxies the
\lz\ relation is explained in the scenario of mass-loss through galactic
winds.  
However, dIrrs are still evolving systems, 
so if we want to maintain the mass-loss scenario, 
then we have to assume 
that it 
leaves behind a \lz\ relation 
very early in the history of galaxies.
Recently  Skillman et al. (\cite{skillman_scl_hii}) and
Pilyugin et al. (\cite{pilyugin_etal_04}) have found that the
chemical evolution of dIrrs can be approximated by a closed-box,
provided that a low  effective yield ($\approx 1/3$ of the standard) is
adopted: this is a typical signature of gas exchange with the
environment.
Pushing this hypothesis further, we plot such closed-box
models\footnote{More appropriately, one should perhaps call these models
``open-boxes simulating low-yield closed-boxes''} in the right panel of
Fig.~\ref{fig:lz}, for total masses $m_{\rm tot}=m_{\rm gas}+m_{\rm
stars}$ varying between $5\times 10^7 ~m_\odot$ and $10^9~m_\odot$.  We
adopted the proposed low yield $p_{\rm (O/H)}=1.6
\times 10^{-4}$, 
and we assumed $M/L=1.2$, $M_{H,\odot}=3.1$ and $(m-M)=28$ for both
groups.  The gas mass fraction $\mu=m_{\rm gas}/m_{\rm tot}$ decreases
monotonically along each track as abundance increases, so if one assumes
that all galaxies were born at the same time, then the figure is telling
us that more massive galaxies evolve faster along their tracks.  This is
confirmed by plotting $d\mu/dt$ vs. total mass from data published in
Skillman et al. (\cite{skillman_sfr}): one can see a general trend of
$d\mu/dt$ increasing with total mass, i.e.  larger galaxies being more
effective in converting their gas into stars. A similar conclusion was
reached by Pilyugin \& Ferrini (\cite{pilyugin_ferrini}), who found that
the \lz\ relation is a combined effect of smaller gas loss and higher
astration level as the mass increases. More detailed modeling is needed
to clarify the role of other processes such like infall
(e.g. Hidalgo-G\'amez et al. \cite{anamaria03}).

It is also interesting to note that, although 
the closed-box 
predicts arbitrarily large
values of $Z$ as 
$\mu$ goes to zero
($Z=p\ln (1/\mu)$; Searle \& Sargent
\cite{ss72}), if we assume that the mass of each new generation
of stars, which we can call $m_{\rm SF}$, is roughly constant, then the
maximum metallicity we can measure will be 
$Z_{\rm max}= p ~ \ln(m_{\rm tot}/m_{\rm SF})=p ~ \ln m_{\rm tot} - Z_0$.
This dependence on  the mass of the galaxy is shown by
the dashed curve in Fig.~\ref{fig:lz}, where we assumed $m_{\rm
SF}=2\times 10^7 m_\odot$, a mass of stars that can be created in $\sim
10^8 \rm yr$ for a typical SFR~$\sim 0.1~m_\odot
\rm yr^{-1}$.

\section{A dwarf  vs. giant galaxy dichotomy?}

\begin{figure}[t]
\centerline{
\includegraphics[width=10.5cm]{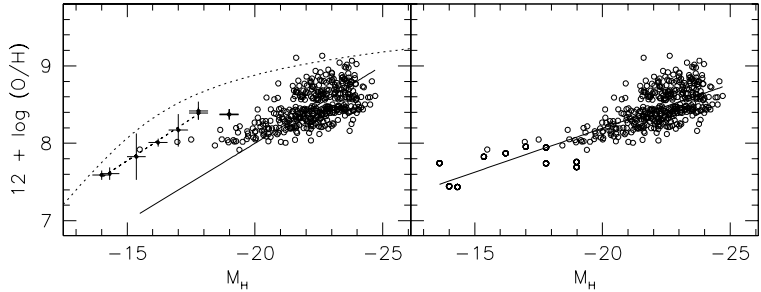} } 
\caption{ 
Oxygen abundances vs. $H$ luminosity for both the S05 emission-line
galaxies (KBG calibration) and our dwarf irregulars. In the left panel
we plot the direct abundances, 
while abundances from the P00 method are used
in the right panel. The left panel shows also the
\lz\ relation that S05 fit to their galaxies, and the fit we obtain for
our dwarfs. The dotted line is the same curve shown in
Fig.~\ref{fig:lz}, and in the right panel the solid line is an
unweighted fit through the whole sample.
\label{fig:salzer}}
\end{figure}

The only study of the \lz\ relation of giant galaxies that includes
near-IR magnitudes is that of Salzer et al. (\cite{salzer05}; hereafter
S05). They took spectra of emission-line galaxies in fixed apertures of
1.5$^{\prime\prime}$ or 2$^{\prime\prime}$, and computed oxygen
abundances with a reduced number of emission lines, due to the limited
spectral coverage of their data:
using additional spectra, they first obtained metallicities for a
subsample of galaxies using both the direct and the $R_{23}$ methods,
and then calibrated them vs.  [\ion{N}{II}]$\lambda$6583/H$\alpha$ and
[\ion{O}{III}]$\lambda$5007/H$\beta$.  
For the $R_{23}$ method they used
the
P00 calibration
for the lower branch of
the $12+ {\rm
\log(O/H)}$ vs. $R_{23}$ relation, and three calibrations for the upper
branch: Edmunds \& Pagel (\cite{edmunds_pagel84}; EP), Kennicutt,
Bresolin, \& Garnett (\cite{kbg03}; KBG), and Tremonti et
al. (\cite{tremonti_etal04}).  Assuming again an average distance
modulus $(m-M)=28$, in Fig.~\ref{fig:salzer} we plot our data together
with Salzer's et al. data.  The figure shows that the scatter in the
\lz\ relation for giants is very large, compared to that of dwarfs,
perhaps due to fixed-aperture effects and uncertainties intrinsic to the
empirical methods (see also the discussion in S05).  The straight lines
in the left panel are the fits obtained by S05 and by us, and taken at
face value, the panel would suggest a well-defined offset between the
\lz\ relation of giants and dwarfs. The slope of the S05
\lz\ relation is in fact close to what we find for dwarfs alone, namely 
$-0.215\pm0.003$ and $-0.201\pm0.004$ for abundances obtained with the
EP and KBG calibrations, respectively.  Now to be consistent with S05,
in the right panel we plot our abundances computed with the P00 method
(which allows to add AM~106-382). The offset between galaxies in the two
mass ranges seems to disappear: it is in fact possible to do a linear
fit of the whole sample (solid line), although a change of slope at
$M_H\approx -20$ would seem a better choice.  The question of a
dwarf-giant dichotomy is then open: moving from the empirical to the
direct abundances, the \lz\ relation for dwarfs becomes much better
defined, parallel to the one for larger galaxies, and with a substantial
offset. It remains to be seen what would happen to the \lz\ relation of
giant galaxies: we plan to measure direct abundances for some of the S05
galaxies, to see if the offset is confirmed or not.  If the scenario of
the top panel were confirmed, then we should conclude that dwarf
galaxies are able to make more metals than giants, which at first sight
is puzzling.  A possible explanation is the following: if we assumed
that the fundamental relation is that of dIrr galaxies, then we could
say that the amount of metals produced by a giant galaxy is the same as
that produced by a dwarf galaxy ca. $10$ times less luminous
($\sim2.7$~mag). In other words, only $10\%$ of the disk of a giant
galaxy would participate in making its metals, or perhaps only $1\%$ if
we considered that the yield in giant galaxies could be some 3 times
higher than in dwarfs.  Metaphorically, one could think that metals in
the central regions of star-forming galaxies are made in universal
cells, and that large cells make metals faster than small cells. And
while small cells are found in isolation (dwarf galaxies), larger cells
are found in groups of $10$--$100$ in the centers of giant galaxies.
This argument is certainly appealing, but it needs confirmation with a
follow-up of the S05 study. In fact at the moment we can somewhat
reconcile the two \lz\ relation only at the price of a degradation of
our data. It is finally worth mentioning that, since the S05 sample
includes star-forming galaxies, one could suspect that giant galaxies
are offset in luminosity due to the presence of a star-burst: however,
Lee et al. (\cite{lee_etal04}) find that this offset is a few tenths of
magnitude in $B$, compared to quiescent galaxies. We expect that the
effect in the IR must be even lower, and certainly not comparable to the
one we observe.

\begin{acknowledgments}
We thank John Salzer for useful discussions and for providing his data
in electronic format, and Jay Gallagher for important remarks in the
course of the meeting. I.S. wishes to thank Ana Maria Hidalgo-G\'amez
for her kind invitation to UNAM, where  some issues of this work
were discussed.
\end{acknowledgments}

\end{document}